# Light Yield Quenching and Quenching Remediation in Liquid Scintillator Detectors

S. Hans,[a,b] J.B. Cumming,[a] R. Rosero,[a] R. Diaz Perez,[a] C. Camilo Reyes,[a] S.S. Gokhale,[a] M. Yeh[a,c,*]

[a]*Chemistry Division, Brookhaven National Laboratory, Upton, New York 11973*
[b]*Chemistry Department, Bronx Community College, Bronx, New York 10453*
[c]*Instrumentation Division, Brookhaven National Laboratory, Upton, New York 11973*
E-mail: yeh@bnl.gov

ABSTRACT: Quenching of light emission from an LAB based scintillator by the addition of organic amines and carboxylic acids is examined. Chemical functional groups of the quenching agents play an important role in this reduction. It is shown that "salt" formation at a 1:1 mole ratio in a mixed amine-acid system, reduces quenching by a factor of ∼ 2. Supporting NMR spectra are presented. This "quenching neutralization" has the potential to reduce the light loss incurred when metals complexed with quenching agents are loaded into organic scintillators.

[*]Corresponding author.

**Contents**

**1. Introduction**

Over the past decades, organic liquid scintillators (LSs) have played an important role in many physics experiments; from the first detection of reactor neutrinos by Reines and Cowan [1]; in current measurements of the neutrino mixing parameter, $\theta_{13}$, by the Daya Bay [2], RENO [3] and Double-CHOOZ [4] experiments; in double-beta decay studies [5, 6]; and in the proposed LZ dark matter search [7]. A generic liquid scintillator consists of two basic components; a solvent and a fluor. An additional fluor (wavelength shifter) is often added to redshift the primary-fluor light for better overlap with the sensitivity curve of a photomultiplier tube (PMT). A frequently used fluor is 2,5-diphenyl-oxazole (PPO) with 1,4-bis-(o-methylstyryl)-benzene (bis-MSB) as the wavelength shifter. We adopt linear alkyl benzene (LAB) containing 3 g/L PPO and 15 mg/L bis-MSB as our reference standard (rLS). For many applications, a metal is also added to the LS, e.g., gadolinium (Gd) to detect neutrons produced in the reaction: $\bar{v}_e + p \rightarrow e^+ + n$, or tellurium (Te) as the source of double-beta decay electrons. For a general review of liquid scintillators and metal loading, see Buck and Yeh [8].

      An electron loses energy primarily by ionization when passing through matter. Information on electron stopping in various elements, compounds, and mixtures is available from the NIST program ESTAR [9]. For LAB, the specific energy loss, *dE/dx*, increases from 1.90 MeV cm$^2$/g at 1 MeV to 23.8 MeV cm$^2$/g at 10 keV. The range, $R = \int_0^E (dE/dx)^{-1}\, dE$, decreases from 0.42 g/cm$^2$ to $2.4\times10^{-4}$ g/cm$^2$ over that region. For PPO, *dE/dx*, varies from 1.97 MeV cm$^2$/g to 21.5 MeV cm$^2$/g over the same range.

      The electron track is an excited region containing charged molecular fragments, free radicals, and secondary electrons. Some of the energy will be in the form of electronically and vibrationally excited LAB molecules. If their energy is below the ionization energy, IE = 9.1 ± 0.1 eV for dodecylbenzene [10], a cascade of internal conversion and vibrational relaxation will follow. Some energy may be transferred to other LAB or PPO molecules, but most dissipates as heat. Some LAB molecules reach the $S_1(0)$ level from which fluorescence occurs. ($S_n(v)$ represents $v_{th}$ vibrational member of the $n_{th}$ singlet electronic band.) For a more detailed discussion than the following, see Ogawa, Yamaki, and Dato [11].

      Presence of an aromatic ring in the solvent is conducive to high light yield (LY). Consider benzene as the simplest aromatic chromophore. A chromophore is that part of a molecule responsible for its light absorption and emission. In benzene, sp$^2$ hybridization results in six equivalent C-C bonds and a symmetric planar core structure (bond order 1.5). The remaining p orbitals hybridize to form six delocalized π molecular orbitals. Filling the lowest three with six ″valence″ electrons results in a singlet $S_0(0)$ ground state of π character. The optical absorption $S_0(0) \rightarrow S_1(2)$ is a $\pi \rightarrow \pi^*$ transition.

Döering [12] has proposed a level structure for benzene based on his inelastic electron scattering measurements and the work of others. Levels up to 7 eV are classified into singlet (S) or triplet (T) systems. Intersystem crossings such as the phosphorescence transition, $T_1(0) \to S_0(0)$, do not play a significant role in scintillator light production. They are hindered by factors of the order of 10.

Recent optical absorption and emission spectra [13] complement the work of Döering. When plotted as a function of $v$, the energies of $S_1(0)$ to $S_1(8)$ define a line having slope $0.098 \pm 0.001$ eV and an intercept at 4.64 eV. The latter is our best estimate for $S_1(0)$, the band gap of the benzene chromophore. Döering has noted that constant vibrational spacing is characteristic of both singlet and triplet bands. He ascribes his value for the spacing, $0.107 \pm 0.009$ eV, to the symmetric breathing mode vibration of the benzene ring. The optical emission spectrum of benzene approximates a low-resolution reflection of the absorption one. Highest probabilities occur for the $S_1(0) \to S_0(2)$ emission and $S_0(0) \to S_1(2)$ absorption transitions. Difference between absorption and emission maxima places the *Stokes' shift* at 0.54 eV. This energy loss on an up-down cycle corresponds closely to five vibrational quanta.

The level schemes shown in Fig. 1 for LAB as a donor chromophore and for PPO as an acceptor, were derived from tabulated values of the absorption and emission maxima [8] and the assumption that the *Stokes' shift* corresponds to five vibrational quanta as was observed for benzene. The down arrow at the upper left indicates events feeding into the $S_1$ band of LAB from more highly excited states. (Some direct feeding of that band of PPO may also occur.) Rapid vibrational relaxation (orange dashed arrows) within the bands lead to population of the $S_1(0)$ levels from which fluorescence occurs (bold labeled arrows). These are shown terminating at $S_0(2)$, the most probably transition in benzene. Transitions to $S_0(0)$ are a factor of ~ 3 less probable.

Two modes of energy transfer from LAB to PPO are shown in the figure. Förster Resonance Energy Transfer [14] (FRET) is indicated by the red horizontal arrow connecting $S_1(0)$ of LAB to $S_1(9)$ of PPO. This involves dipole-dipole interactions between the states, with the exchange of a virtual photon. In its original form, FRET had an $r^{-6}$ dependence on donor-acceptor distance. Recent work indicated longer range ($r^{-4}$ and $r^{-2}$) components, see the review by Jones and Bradshaw [15]. Energy can also be transferred to the PPO by reabsorption of a real LAB fluorescence photon, the blue wavy line labeled emission. When absorbed by the acceptor or another nearby PPO molecule it promotes an electron to the $S_1(7)$ level. Regardless of the transfer mechanism, light is emitted at the PPO fluorescence wavelength.

An overall picture of energy propagation through the LS is shown in Fig 2. The ordinate here is the light wavelength. Pairs of horizontal lines indicate tabulated maximum absorption and emission values [8] for each of the three scintillator components, LAB, PPO, and bis-MSB. The response curve of a typical blue-green sensitive PMT is shown on the right ordinate. Starting from the lower left, the deexcitation cascade feeds LAB, from which fluorescence light (peak $\lambda$ = 284 nm) may be emitted. Were this to reach the PMT (the horizontal wavy arrow), it would be near the lower limit of sensitivity. Pure LAB has a LY of ~ 7% of that attainable by optimizing the PPO concentration [16].

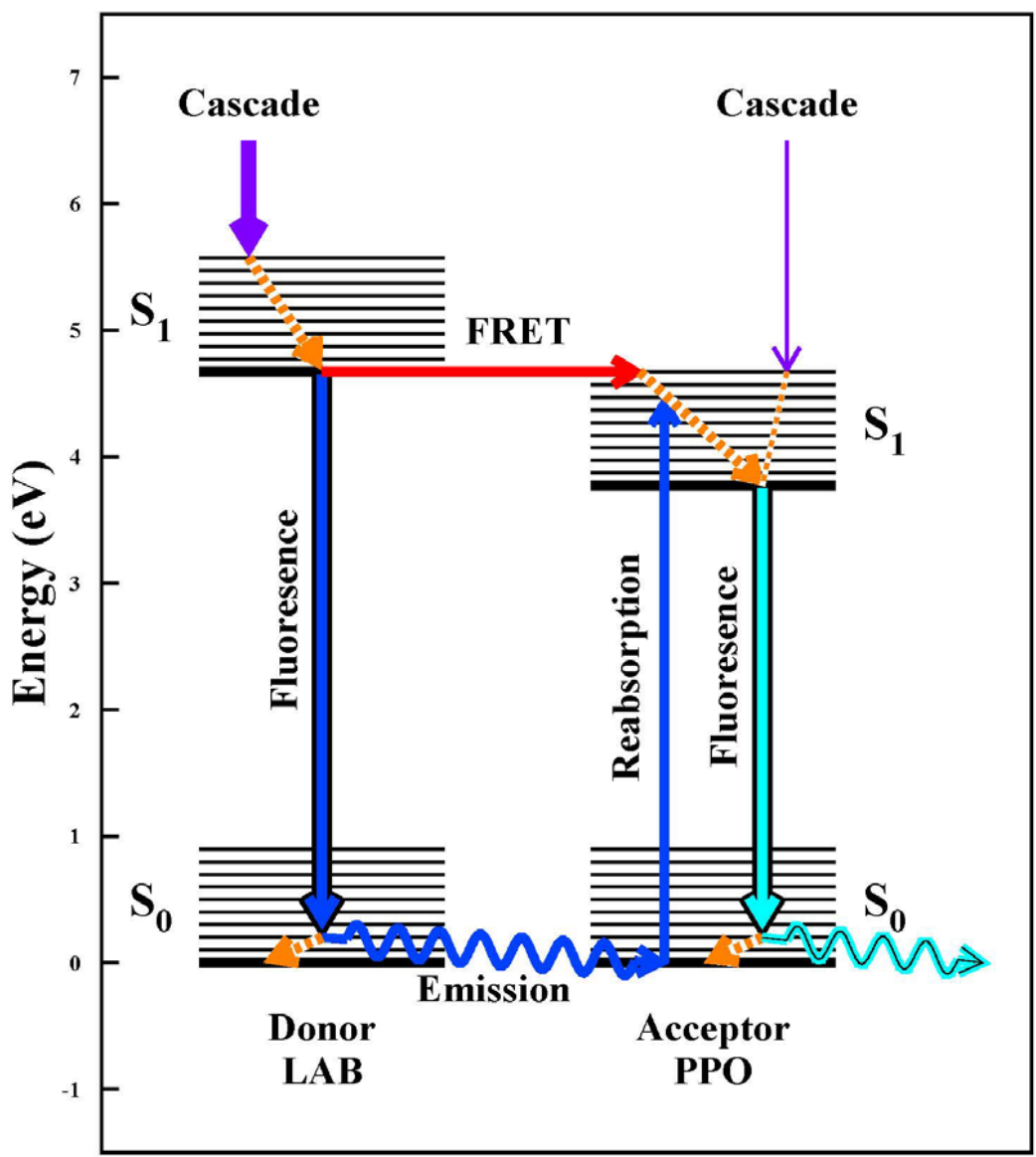

**Figure 1**. Schematic diagram of low-lying levels of LAB and PPO showing paths for energy transfer. Arrows at the top represent energy input from the initial excitation. Diagonal dashed arrows (orange in color) indicate rapid vibrational relaxation. Fluorescence light emission from the $S_1(0)$ states of LAB and PPO is shown by bold down-arrows (blue and cyan in color, respectively). Two modes transfer energy from LAB to PPO: the one shown by the horizontal red arrow labeled FRET, involves dipole-dipole interactions and transfer of a virtual photon between the chromophores, the Förster Resonance Energy Transfer mechanism. The other, emission and reabsorption of a real photon, is illustrated by the wavy blue line.

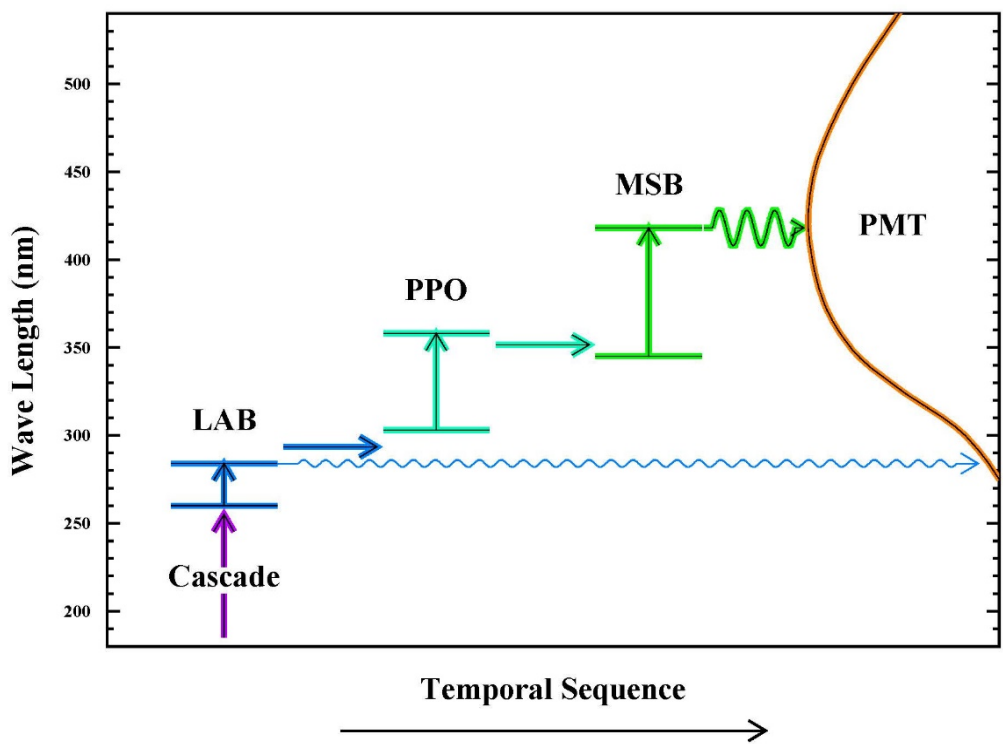

**Figure 2.** Progressive red shifting (*Stokes*) of light in an LAB based liquid scintillator. Two horizontal lines indicate experimental maximum absorption and emission wavelengths [8] for each of the components, LAB, PPO, and bis-MSB. The response curve for a typical blue-green sensitive PMT is shown on the right of the figure.

Two horizontal arrows in Fig. 2 show excitation transfer from the LAB to the PPO and from the PPO to the bis-MSB, by either virtual (FRET) or real photons. The maxima of the donor emission and acceptor absorption spectra do not generally match, but the spectra are broad and overlap sufficiently for effective energy transfer. Fluorescence from the bis-MSB (peak $\lambda$ = 418 nm) is close to the maximum PMT sensitivity. On our light yield scale of 0-100, the overall LY for rLS is 95.5.

In the present paper, we examine the reduction of LY, ″quenching″, on the addition of organic amines and carboxylic acids to an LAB based scintillator. These two classes of compounds, together with others such as alcohols, diols, or diketones are frequently used as complexing ligands or chelating agents to introduce metals into an LS for nuclear and particle physics experiments. The concurrent quenching places a practical limit on attainable metal concentrations. It was thought that details of the chemistry of quenching might point to procedures to reduce this effect.

Light quenching can be grouped into three general classes:

<u>Ionization Quenching</u>: When energy deposition, *dE/dx*, becomes high near the end of an electron′s range, recombination on the ions and electrons in the track will reduce the energy available for light production, *dE″/dx*, an additional variable in a light yield study. The Birks′ quenching equation

[17]

$$\frac{dE'}{dx} = \frac{(dE/dx)}{1 + k_B(dE/dx)} \quad (1.1)$$

relates to the energy available for a reaction, $dE'/dx$, to the input energy, $dE/dx$, where $k_B$ is quenching constant. The observed light, $dL/dx$, is given by

$$\frac{dL}{dx} = S\left(\frac{dL}{dx}\right)_{abs}\left(\frac{dE'}{dx}\right) \quad (1.2)$$

where $\left(\frac{dL}{dx}\right)_{abs}$ is the absolute light and $S$ is the efficiency of the detection system.

Many experiments including the present and that of Kögler et al. [18] are integral measurements: An electron is generated in and stopped by the sample, so $L = \int_0^R \left(\frac{dL}{dx}\right)dx$ is to be compared with the input energy, $E^o = \int_0^R \left(\frac{dE}{dx}\right)dx$. Kögler et al. report proportionality between the two for $E^o > 100$ keV; $k_B$ is small. Birks′ quenching can be ignored for the 477-keV electrons used in our measurements.

<u>Color Quenching</u>: When light progresses through a scintillator it may be absorbed by colored organic impurities or inorganics such as $Fe^{3+}$. The reduction of attenuation length by impurity quenching becomes important for large detectors where scales are measured in meters. It can be overcome by careful purification of the scintillator components [8].

*Self-quenching* (*Concentration quenching*) is also important for large detectors. When the emission and absorption spectra of a material overlap, lack of transparency to its emitted radiation reduces the attenuation length. Multiple *Stokes′ shifts*, as shown in Fig. 2, avoid this problem. A special type of color quenching also occurs when a colored metal target (i.e. neodymium, Nd) is loaded into an LS for a double β-decay study. Light absorption by the Nd 4f bands in the region of PMT sensitivity cannot be avoided.

<u>Chemical Quenching</u>: The generalized acid-base formalism of Lewis [19] provides a convenient framework for discussing the chemical effects studied in the present work. In an LS, organic acids and amines can function as electron acceptors and donors, respectively. Separately, they are effective light-yield quenchers. However, quenching is reduced by a factor of ∼ 2 when they combine to form a salt in which the acceptor-donor effect is neutralized. Our results suggest that some reduction of quenching also occurs when two acid molecules combine to form a dimer.

**Experimental**

Linear alkyl benzene (LAB) from Cepsa Canada, was purified using neutral $Al_2O_3$ columns. Butyl amine (99%), octyl amine (99%), dodecyl amine (99%), N,N-dimethyl-dodecyl-amine (97%) and propionic acid (99.5%) were purchased from Sigma-Aldrich. Trimethyl hexanoic acid, (95%) from Chemos GmbH, was purified by thin-film vacuum distillation. Scintillator grade 1,5-diphenyl oxazole (PPO) and 1,4-bis(2-methylstyryl) benzene (bis-MSB) were acquired from Research Products International (RPI). The PPO was purified by recrystallization from methanol. LAB doped with 3 g/L PPO and 15 mg/L bis-MSB served as the reference standard (rLS) for our measurements.

Light yields were measured using procedures described previously [16]. To summarize; a Beckmann LS-6500 Beta Ray Scintillation Spectrometer had been modified to hold a $^{137}$Cs source at a fixed distance from the sample position. Nominal 10 mL samples were contained in standard 20 mL scintillation vials. Oxygen is a known chemical quencher. It was not practical to degas and keep our samples degassed during the analysis procedure. Quenching by air was determined in one measurement of a degassed (by $N_2$ bubbling) LS sample. The resulting increase of 12 ± 1% over standard rLS agrees with the 11% reported by Xiao et al. [20].

Photons from excitation of the sample by Compton electrons are collected by two photomultiplier tubes. A prompt (~ 20 nsec) coincidence between the PMTs triggers digitization and storage of their sum signal. This "two-photon" trigger reduces a variety of background effects. A key feature of the POWder-DERivative (POWDER) data analysis procedure is the conversion of an observed LS-6500 event spectrum, $N(E)$, into a power spectrum, $P(E) = E \times N(E)$. Comparison of the position of the $P(E)$ derivative peak with that of the rLS gives the relative LY. One hundred on the LY scale had been set as the maximum light from the LAB+PPO binary system (PPO concentrations 4.5-5.5 g/L) [16].

Extensive use was made of the automation capabilities of the LS-6500. For a typical measurement, 11 unknowns and an rLS sample were loaded, and the machine instructed to assay the set 12 times. It was thought that averaging would reduce the 1-2% scatter observed previously [16]. Some scatter remained. This appeared related to variations in overall system gain: Dispersion of absolute LY values was greater than those measured relative to the rLS. Accordingly, the mean LY for an unknown was normalized to place rLS in that set at LY= 95.5. We believe that a conservative error estimate for a LY determined this way is $\sigma_{LY} = \sqrt{(0.01 \times LY)^2 + 0.3^2}$.

The two-photon coincidence results in an approximately error function cutoff of event spectra (50% point at channel ~1.5, σ ~0.8 channels). For LY> 0.5, the upper side of the power spectra and its derivative peak show little distortion from the cutoff. For lower LYs, the observed derivative maximum appears to be an upper limit to the true peak position. In those cases, the LY was calculated from the observed value assuming a 100% uncertainty.

Proton NMR spectra were obtained with a Bruker 400-MHz UltraShield Spectrometer using deuterochloroform ($CDCl_3$) as the solvent. The spin ½ of a proton can assume either parallel or antiparallel orientations with respect to the magnetic field, $B^0$, of the Brucker. The energy difference between those states, $\Delta E$, is proportional to the effective field, $B$, at the nucleus which is less than $B^0$ due to shielding by electrons in atomic and molecular orbitals. It is conventional to report peak positions not in terms of $\Delta E$, but as chemical shifts, δ, measured in ppm from a zero set by the singlet peak for tetramethyl silane, $Si(CH_3)_4$. Values of δ are independent of $B^0$. They depend on chemical factors such as the atom to which the proton is attached and the presence of nearby functional groups. A positive δ is referred to as "downfield". The area of a peak is proportional to the number of protons in that environment.

**Results and Discussion**

Acronyms, chemical names, formulas and mass numbers of compounds discussed in this work are listed in Table 1. Chemical formulas are in a form which emphasizes they consist of an alkyl chain (to the left of the long dash) and a functional group (to the right). Linear alkyl benzene, the solvent of rLS, has a range of chain lengths. That for the average composition is shown in the table. LAB is frequently approximated as dodecyl benzene, $C_{18}H_{30}$, A = 246.

**Table 1.** Compounds discussed in this work.

| Acronym | Chemical Name | Formula | Mass Number |
|---------|---------------|---------|-------------|
| LAB | linear alkyl benzene | $C_{11.2}H_{23.4}$—$C_6H_5$ | 234.8 |
| BA | butyl amine | $C_4H_9$—$NH_2$ | 73 |
| OA | octyl amine | $C_8H_{17}$—$NH_2$ | 129 |
| DMDDA | N,N-dimethyl-dodecyl-amine | $C_{12}H_{25}$—$N(CH_3)_2$ | 213 |
| PA | propanoic acid | $C_2H_5$—COOH | 74 |
| TMHA | 3,5,5-trimethyl-hexanoic acid | $C_5H_9(CH_3)_3$—COOH | 158 |

Five light quenching spectra obtained with the LS-6500 and POWDER analysis procedure are shown in Fig. 3. These trace the reduction in light yield as a portion of the LAB in the rLS is replaced by the quenching agent. All samples contain the same concentration of fluor, 3g/L PPO with 15-mg/L bis-MSB. Because the fluor was added as a concentrate in LAB, the maximum concentration of amine or acid quencher was 97% by weight. The mole fraction abscissa emphasizes the importance of functional groups in the quenching process. Were the quenching materials identical to LAB in all respects except having zero light yield, measured LYs would fall linearly as shown by the dotted lines in the figure.

Data for three amine quenchers shown in Fig. 3(a) indicate that quenching by OA is identical to that by BA on a mole to mole basis. The four-carbon increase in alkyl chain length has an insignificant effect. Further increasing the chain to twelve carbons in DMDDA does lead to significantly greater quenching. Note, however, that DMDDA is a tertiary amine, while both BA and OA are primary amines.

These observations can be described qualitatively in terms of Lewis′ generalized theory of acids and bases as electron acceptors or donors, and Pauling's concept of bond polarization and electronegativity [21]. The lone pair of unbonded electrons on nitrogen make ammonia, $NH_3$, a Lewis base. The methyl group (-$CH_3$) and alkyl groups in general are electron donors. Substituting one for a proton of NH3 adds an incremental negative charge to the nitrogen and increases basicity. Additional groups to form secondary and tertiary amines magnify the effect. These electronic inductive effects place amine basicity in the order

$$NH_3 < primary < secondary < tertiary \qquad (3.1)$$

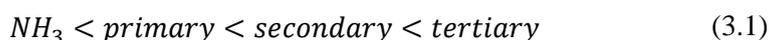

Experimental measurements indicate that other factors play a role. Graton et al. [22] quantify basicity in terms of the strength of the hydrogen bond between a test amine and 4-fluorophenol (basicity = $-\Delta H^o$ kJ/mole). Values for 68 amines and ammonia span a wide range of basicity, which ″is mainly explained by the basicity-enhancing electronic effects of alkyl groups, which can be overcompensated by dramatic basicity-decreasing steric effects.″

In an attempt to separate the two effects, we extracted from their database basicities for acyclic amines having alkyl substituents no larger than butyl. Mean basicity values for ammonia and the selected primary, secondary and tertiary amines were; 31.5. 34.0, 35.3 and 36.2, respectively, the direction expected from the inductive effect alone. However, the extracted tertiary distribution is bimodal. Triethylamine (39.4) and tri-n-butylamine (38.4), the upper group, are the most basic amines of all those studied by Graton et al. They suggest a strong continuation of the inductive effect. The low group mean is 34.1±0.5, identical with that of the primary amines. That group comprises N,N-dimethyl-isopropyl-amine (34.7), N,N-dimethyl-ethyl-amine (34.5), N,N-diisopropyl-ethyl-amine (33.9) and trimethyl-amine (33.5). It is unreasonable to account for the difference between groups as a steric effect as tri-n-butylamine and triethylamine should be more encumbered than trimethyl amine.

A variety of data indicate base strength is independent of n-alkyl chain length for primary amines, hence, the identical light quenching observed for butyl- and octylamine might be expected. This establishes one point on a possible quenching-basicity correlation curve. The more-highly light-quenching amine, N,N-dimethyl-dodecyl-amine (DMDDA) is a member of the homologous series: N,N-dimethyl-hydrogen-amine (35.1), N,N-dimethyl-methyl-amine (33.5), and N,N-dimethyl-ethyl-amine (34.5). This emphasizes the lack of a basicity increase between the secondary and the two tertiary amines. Considering that any steric effect would be negative, it seems unlikely the basicity of N,N-dimethyl-dodecyl-amine would be greater than that of its lower homologues. However, absent a direct measurement and considering the bimodal distribution for tertiary amines, no convincing quenching-basicity correlation should be inferred from the present data.

Light quenching by two carboxylic acids, PA and TMHA, is compared in Fig. 3(b). TMHA is the organic acid used to introduce Gd into LS [23]. While differences are larger than those between the primary amines, the general shape is the same for both acids. In terms of the electron donating effect on the -COOH, once past methyl (-$CH_3$), the difference between -$C_2H_5$ and -$C_5H_8(CH_3)_3$ should be small.

There is a qualitative difference between the amine and carboxylic acid quenching curves. Light yields for both decrease rapidly with increasing concentration for mole fractions < 0.15. This trend continues to higher concentrations for the amines, but curves for the acids transition into a slower, nearly linear decrease. This may reflect dimer formation in the carboxylic acids as discussed below.

Early observations had suggested that adding one quenching compound to another in a scintillator did not have an additive effect: In some cases light yields increased. The present work examined this effect using a mixed quenching system of octyl amine (OA) and propanoic acid (PA), as a constant 15% by weight of the LAB in rLS was replaced by OA-PA mixtures. Fluor concentrations in the samples were maintained at the standard 3 g/L PPO plus 15 mg/L bis-MSB. The dependence of light yield on mole ratio, R, of the added PA/OA is shown in Fig. 4. Light yields rise by a factor of ~2 starting from 15% OA in LAB on the left, to a peak at R ~ 1, then fall up to R = 2. The decrease must continue, eventually reaching (for R $\Rightarrow \infty$) the LY for 15% PA in LAB shown by the arrow in the figure. In the absence of PA-OA interactions, the LY would vary along a smooth curve between those limiting values.

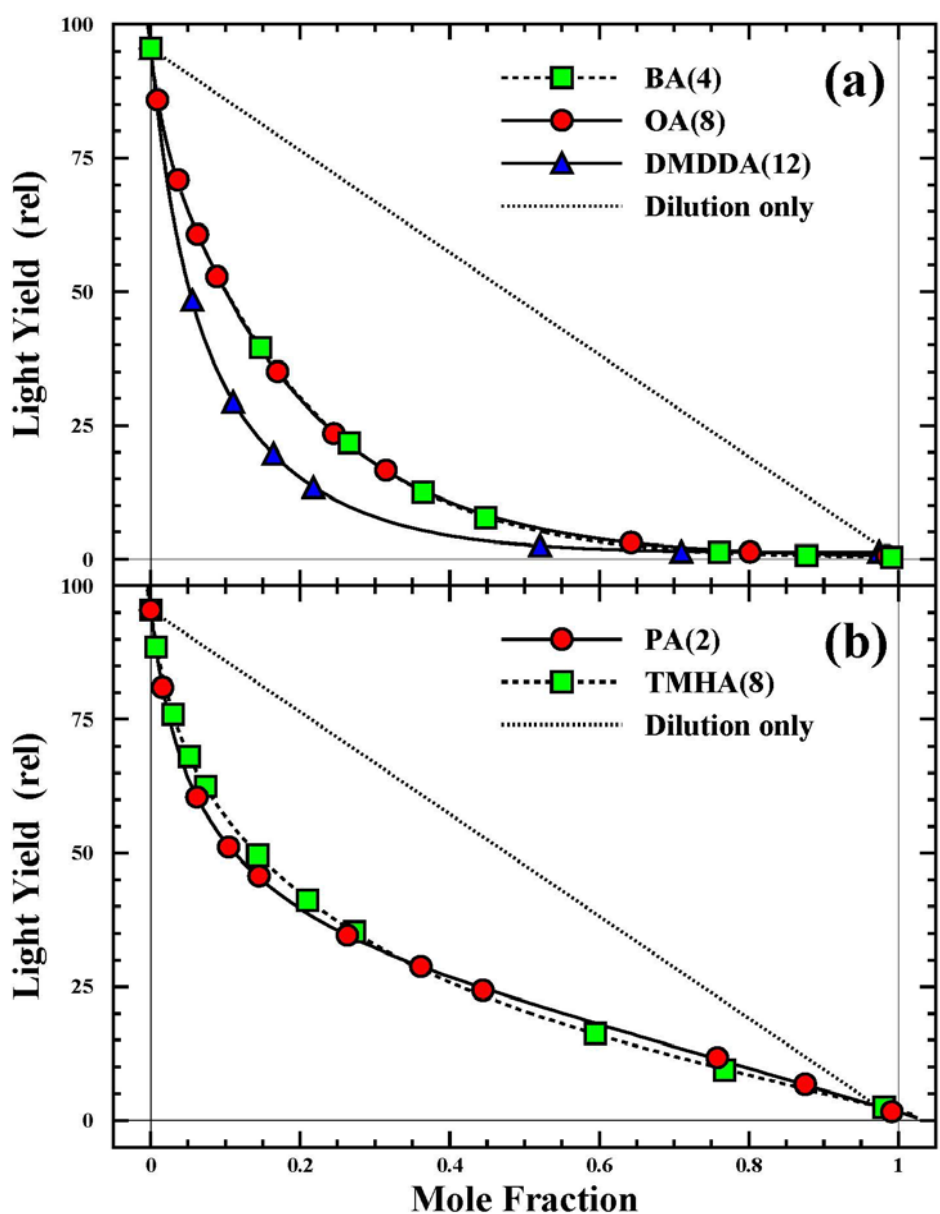

**Figure 3**. Light quenching spectra for three amines in (a) and two carboxylic acids in (b). These show the decrease in light yield when LAB in the rLS is replaced by the indicated light quenching compound. The mole fraction abscissa emphasizes the relationship between chemical functional group and light quenching. Smooth curves indicate general trends. The number in parentheses following an acronym is the number of carbon atoms in its alkyl chain. Dotted lines show the linear decrease expected if dilution was the only effect.

Intersecting lines in the figure are from a four parameter least-squares-fit(LSF) taking the coordinates of the intersection point and the two slopes as variables. The intersection is at R = 1.013 ± 0.009 with $x^2/ndf$ = 0.20. Although the group of six points directly under the vertex (not included in the fit) suggests some rounding of the top, the data constrain the location of the transition from up-slope to down-slope to 0.95 ≤ R ≤ 1.10.

Insight can be gained by discussing Fig. 4 as a base-acid titration. A conventional titration in aqueous solution uses an indicator sensitive to hydrogen ion concentration, [$H^+$], to determine the equivalence point, R = 1. Starting with pure base (low [$H^+$]), adding acid has little effect initially as it is neutralized to form the salt. In the vicinity of R = 1, [$H^+$] rises rapidly to the high value for the pure acid as all the available base has been neutralized. There is an inflection point at R = 1 and the indicator is triggered. Additional acid has little effect.

In contrast, the light-yield titration in Fig. 4 uses the higher LY of the salt as the indicator. Salt concentration should rise nearly linearly with acid addition up to R = 1, then fall with the additional acid acting as a diluent. A maximum is expected at R = 1 as is seen in the figure. Concentrations of base, acid and salt were calculated as a function of R and the equilibrium constant K for the bimolecular reaction base + acid ↔ salt. Salt dominates for large K when R ~ 1. The predicted dependence of LY on R is shown by curves in Fig. 4 for K = 10 and 100. The LYs were obtained by normalizing the salt concentration for K = 10000 and R = 1 to the LY at the intersection point. The observed sharp change in slope suggests K > 100. The lower light-yield OA and PA have effectively been converted to the OAPA salt which has a factor of two higher LY than either.

Proton NMR spectra provide information on structural changes during this neutralization. Relevant portions of NMR spectra for OA, PA and the OAPA salt are presented in Fig. 5. The abscissa width, $\Delta\delta$ = 2.2 ppm, is the same in the five sections to facilitate comparisons of line width and splitting. Vertical scales are adjusted for display purposes. Sections 5(a) through (c) span a region, which includes the alkyl and amine hydrogens. In each case, consider the functional group, the carboxylic acid (-COOH) or the amine (-$NH_2$), to be on the right. Conventional notation denotes carbon atoms as C1, C2, C3, etc. starting from that group.

Peaks often appear in NMR spectra as ($n+1$)-fold multiplets due to spin-spin interactions with the $n$ protons on adjacent carbon atoms. Multiplet intensity ratios are given by the binomial coefficients, 1:2:1, 1:3:3:1, 1:4:6:4:1, etc. Propanoic acid has a simple spectrum: The two peaks in Fig. 5(a) have areas in the ratio 3:2 suggesting five protons are involved. The 1:2:1 intensity ratio of the triplet at $\delta$ = 1.17 ppm indicates that it is the terminal -$CH_3$ (C2). The 1:3:3:1 pattern assigns the quartet at $\delta$ = 2.40 ppm to the C1 methylene (-$CH_2$-).

The acid proton appears far downfield in Fig. 3(d) as a broad peak at $\delta$ = 11.45 ppm, a value typical of hydrogen bonds. The peak width confirms this assignment. The mean life, $\Delta t$, for the spin relaxation process is related to the intrinsic Lorentz FWHM, $\Delta v$, of the line by the uncertainty principle, $\Delta t \times \Delta v \geq 159$ msec.Hz. The peak at $\delta$ = 11.45 ppm has a Lorentzian shape with $\Delta v$ = 156 Hz, corresponding to a spin relaxation time of 1 msec. The large $\delta$ suggests that the proton is outside the normal molecular framework, i.e., it is inter-molecular. The short time suggests the transient nature of the hydrogen bond between two PA molecules. Propanoic acid is known to form dimers [24, 25], with the most stable ($\Delta H$ = 65 kJ/mole) being the two-bonded ring structure [26]. Dimer formation should partially neutralize the electron donation/acceptance ability of PA, and it might be responsible for the difference in light quenching between PA and OA at high concentrations (Fig.3).

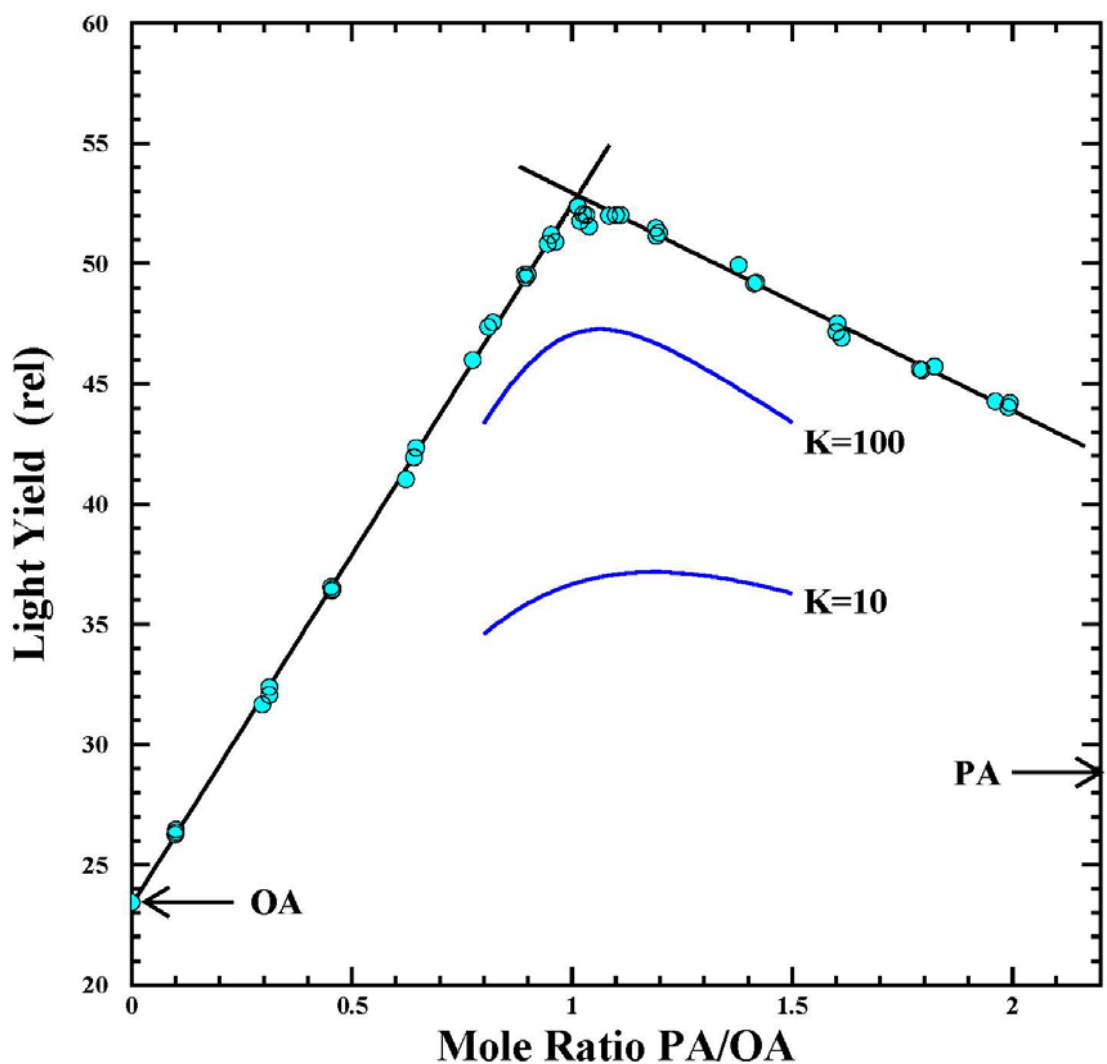

**Figure 4**. Light yield titration curve: Points show LYs observed as 15% by weight of the LAB in the rLS is replaced by propanoic acid (PA) and octal amine (OA) in various mole ratios. The intersecting lines are a LSF to the data, excluding six points at the peak. Curves labeled = 100 and = 10 are calculate salt concentrations, see text. Arrows on left and right ordinates indicate light yields for 15% by weight the pure compounds. Absent salt formation, the LY would vary along a smooth curve connection the two.

The NMR spectrum of octyl amine (OA) in Fig. 5(c) contains 5 peaks. Comparison of the C1 peak of OA at 2.65 ppm, with that from PA at 2.40 ppm, indicates that -$NH_2$ is a somewhat more effective in deshielding than -COOH. The quintet at 1.40 ppm represents the two C2 protons of OA split by the four on C1 and C3. The intense unresolved peak at 1.26 ppm contains the ten C3 to C7 protons. The terminal -$CH_3$ group is at 0.84 ppm.

The singlet peak at 1.08 ppm containing two protons is particularly relevant to the present discussion. It is in the range expected for amine protons. The small value of $\delta$ reflects shielding by the lone-pair electrons on the nitrogen. The assignment was confirmed by adding a drop of $D_2O$. The peak disappeared as labile -$NH_2$ protons were exchanged for deuterium. The peak has a Lorentzian shape with $\Delta v = 7.8$ Hz, corresponding to $\Delta t = 20$ msec. Neither the NMR spectrum of OA nor that of DMDDA (not shown) had peaks indicating hydrogen bonding in these amines.

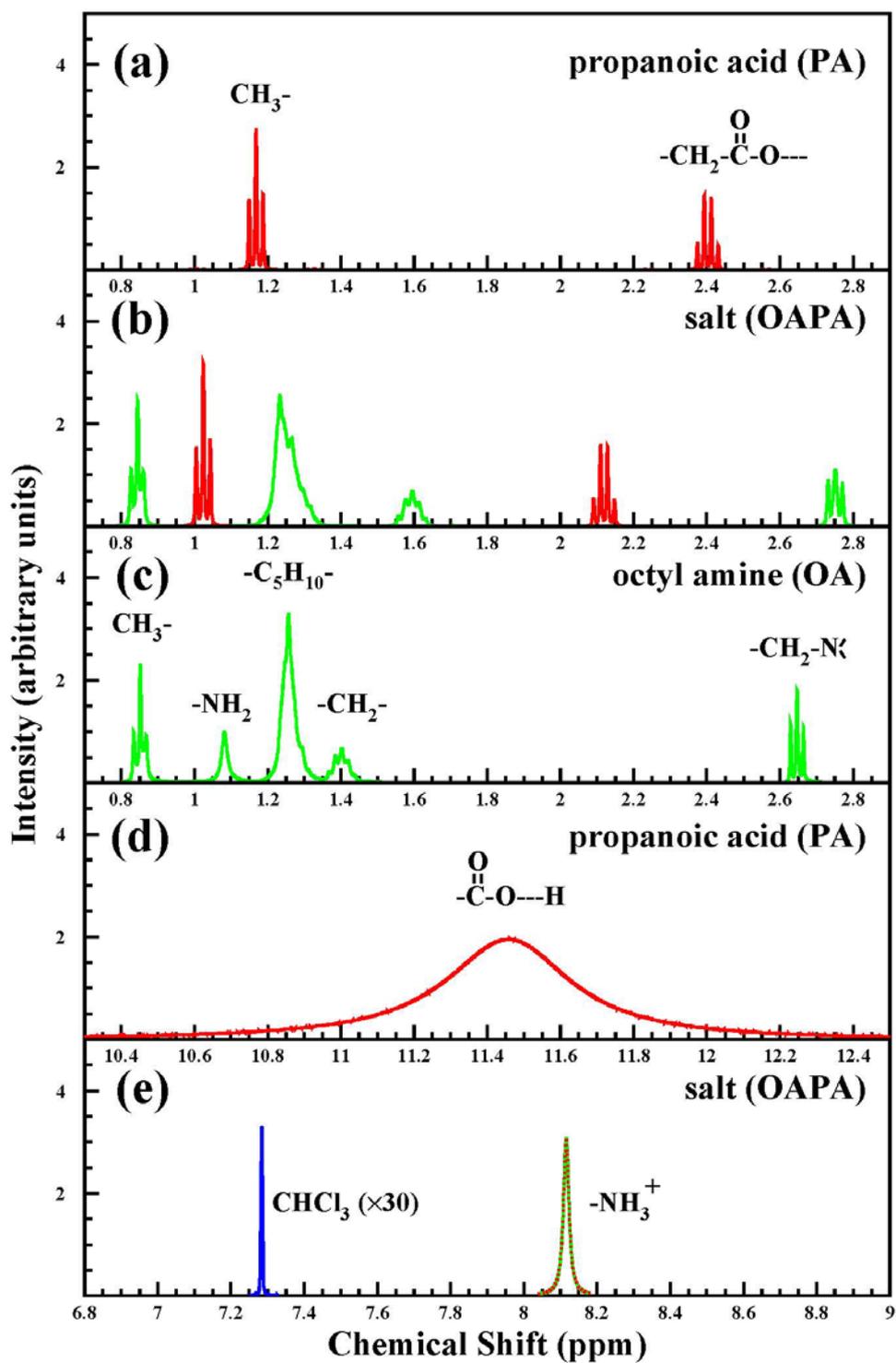

**Figure 5**. Relevant portions of NMR spectra. Intensity (arbitrary units) is shown as a function of chemical shift, δ, for PA, OA and salt (OAPA) samples in deuterochloroform solutions.

The alkyl-hydrogen region of the salt spectrum is shown in Fig. 5(b). Two of the six peaks in the figure trace to PA, four to OA. Intensity ratios confirmed the 1:1 mole ratio of PA:OA in the salt. Absence of peaks from free OA and PA confirm that K for neutralization reaction is large. The two amine protons from PA do not appear in Fig. 5(b), and no hydrogen-bond peak is observed between 9 and 14 ppm. These three missing protons appear in the salt as the broadened singlet at $\delta = 8.12$ ppm in Fig. 5(e). This is a region characteristic of alkyl ammonium cations. The large downfield shift, $\Delta\delta = 7.04$ ppm, is due to the added positive charge on the nitrogen. The line has a Lorentzian shape with $\Delta v = 7$ Hz; spin relaxation time of 23 msec, comparable to that for amine protons, much longer than the 1 msec for the PA hydrogen bond.

The narrow singlet peak at 7.28 ppm in Fig. 5(e) is due to a trace of $CHCl_3$ in the solvent. This provides reference values for a C-H single bond: $\Delta v = 1.7$ Hz; $\Delta t = 95$ msec. The line profile is neither Lorentzian nor Gaussian in shape, suggesting that the natural width for a C-H bond approaches the Brucker resolution.

These NMR data indicate that the neutralization process observed in Fig. 4 involves the transfer of a proton from PA to OA and that Coulomb forces rather than a hydrogen bond hold the anion and cation together in the liquid scintillator. Shifts of $\delta$ for the C1 and C2 protons in the salt from those of the reactants confirm this conclusion. The full negative charge on the $-CO_2^-$ anion in the salt shifts its protons upfield (more shielding). The positive charge on the $-NH_4^+$ cation has the reverse effect, however is much attenuated beyond C2. The chemical shift of the C8 protons is the same for OA and the salt.

A complete light quenching curve was measured for the PA:OA one-to-one complex. Results for the salt are compared with those for the separate compounds in Fig. 6. Only the lower half of the mole fraction range is displayed. This is the region of most practical interest. The logarithmic abscissa permits better comparisons of shapes and differences. Light quenching by -COOH and $-NH_2$ functional groups (in PA and OA) is nearly identical for mole fractions < 0.1. Divergence of the curves at higher concentrations may reflect contributions from less-quenching dimers in the case of PA as suggested by the NMR data.

The increased light yield and reduced quenching for the 1:1 complex observed at one concentration, is seen to be a more general feature extending down to relatively low mole fractions in Fig. 6. Some numerical values of the quenching factor, Q, are presented in Table 2. Q is defined as the percentage reduction of an observed light yield from that of the rLS (LY= 95.5).

**Table 2.** Quenching factors for solutions of propionic acid (PA), octyl amine (OA) and PA/OA=1 mole ratio, in linear alkyl benzene for several total mole fractions (MF) of quenching agent. All solutions contain 3g/L PPO and 15mg/L bis-MSB.

| Compound | MF=0.05 | MF=0.10 | MF=0.15 | MF=0.20 |
|---|---|---|---|---|
| PA | 32.9(7) | 45.4(6) | 53.1(6) | 58.5(5) |
| PA/OA=1 | 16.2(9) | 24.9(8) | 31.7(8) | 37.6(7) |
| OA | 31.8(8) | 47.6(6) | 59.5(5) | 68.7(4) |

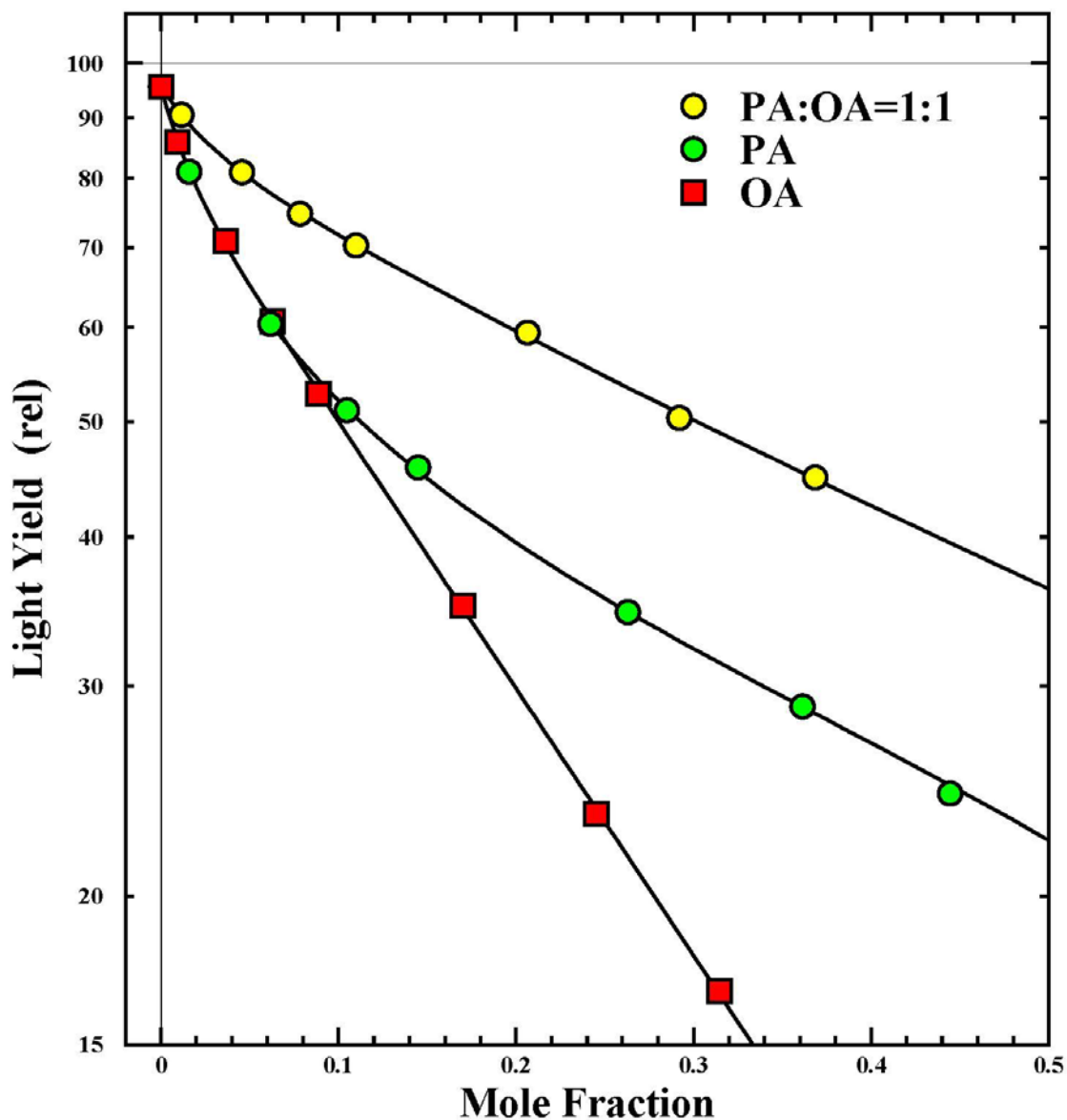

**Figure 6**. Decrease in light yield as a portion of LAB in the rLS is replaced by the indicated mole fraction of PA, OA or PA+OA in a one-to-one mole ratio. Smooth curves show general trends.

**Conclusions**

The reduction of light yield (quenching) when alkyl amines or carboxylic acids are added to an LAB based scintillator is determined largely by the functional group. Quenching is insensitive to details of the alkyl group such as chain length because the electron-donating effect is dominated by the first carbon of the alkyl chain.

Both electron-donating amine bases and electron-accepting acids act separately as quenchers. Primary amines and carboxylic acids exhibit similar quenching power at concentrations < 10 mole percent. The acids become less efficient above that level, possibly due to the influence of hydrogen bonded dimers. A tertiary amine is a stronger quencher than a primary one. There are problems establishing a correlation with experimental amine basicities, especially for tertiary amines. These may involve competition between inductive and steric effects.

The observation of "quenching neutralization" when acid is added to an amine, or vice versa parallels the chemical acid-base reaction to form a salt. NMR measurements indicate that a proton is transferred from the acid to the base during salt formation. The resulting positively charged alkyl ammonium cation and negatively charged alkyl carboxylate anion are held together by Coulomb forces in liquid scintillator. Electron donating and accepting tendencies cancel in the salt which has twice the LY of either base or acid.

Extensions of LY measurements are clearly desirable to establish whether the general trends observed in the present work are applicable to other classes of quenching agents such as alcohols, diols, and ketones. Quenching neutralization has the potential to improve metal loading into LS. It is important then, to study an actual donor-acceptor-metal ternary system to see if higher loading can actually be achieved with reduced quenching, also, whether such a metal-loaded scintillator would have requisite long-term stability by minimizing free-electron mobility for use in large-scale neutrino or other rare-event physics experiments.

## Acknowledgments

The work conducted at Brookhaven National Laboratory was supported by the U.S. Department of Energy under Contract No. DE-SC0012704. The material was based upon work supported by the U.S. Department of Energy, National Nuclear Security Administration, Office of Defense Nuclear Nonproliferation Research and Development.